\documentclass{aa}  
\usepackage{graphicx}
\def\kms    {\ifmmode{{\rm ~km\,s}^{-1}}\else{~km\,s$^{-1}$}\fi}
\def\arcm   {$^{\prime}$}
\def\arcs   {$^{\prime\prime}$}
\usepackage{txfonts}
\usepackage[]{natbib}
\bibpunct{(}{)}{;}{a}{}{,} 
\begin{document}
   \title{The AMIGA sample of isolated galaxies}

   \subtitle{V. Quantification of the isolation\thanks{Full Tables~\ref{tab:ktg}, \ref{tab:hcg} and \ref{tab:crit}--\ref{tab:critHCG} are available in electronic form at the CDS via anonymous ftp to {\tt cdsarc.u-strasbg.fr (130.79.128.5)} or via {\tt http://cdsweb.u-strasbg.fr/cgi-bin/qcat?J/A+A/vvv/ppp} and from {\tt http://www.iaa.es/AMIGA.html}.}}

\titlerunning{The AMIGA sample of isolated galaxies. V.}

   \author{S. Verley
	\inst{1,2,3}
	\and
	S. Leon
	\inst{4,2}
	\and
	L. Verdes-Montenegro
	\inst{2}
	\and
	F. Combes
	\inst{1}
	\and
	J. Sabater
	\inst{2}
	\and
	J. Sulentic
	\inst{5}
	\and
	G. Bergond
	\inst{2}
	\and
	D. Espada
	\inst{2}
	\and
	E. Garc\'ia
	\inst{2}
	\and
	U. Lisenfeld
	\inst{6}
	\and
	S. C. Odewahn
	\inst{7}
        }


   \institute{LERMA -- Observatoire de Paris, 61, avenue de l'Observatoire, 75014 Paris, France\\
              \email{Simon.Verley, Francoise.Combes@obspm.fr}
        \and
             Instituto de Astrof\'isica de Andaluc\'ia -- CSIC, C/ Camino Bajo de Huetor, 50, 18008 Granada, Spain\\
             \email{simon, lourdes, jsm, gilles, daniel, garcia@iaa.es}
	\and
	     INAF -- Osservatorio Astrofisico di Arcetri, Largo E. Fermi, 5 I-50125 Firenze, Italy\\
	     \email{simon@arcetri.astro.it}
	\and
	     Instituto de RadioAstronom\'ia Milim\'etrica, Avenida Divina Pastora, 7, N\'ucleo Central, E 18012 Granada, Spain\\
	     \email{leon@iram.es}
	\and
	     Department of Astronomy, University of Alabama, Tuscaloosa, USA\\
	     \email{giacomo@merlot.astr.ua.edu}
	\and
	     Departamento de F\'\i sica Te\'orica y del Cosmos, Facultad de Ciencias, Universidad de Granada, Spain\\
	     \email{ute@ugr.es}
	\and
	     McDonald Observatory, University of Texas, USA\\
	     \email{sco@astro.as.utexas.edu}
             }

   \date{Received; accepted}

  \abstract
  %
   {The AMIGA project aims to build a well defined and statistically significant reference sample of isolated galaxies in order to estimate the environmental effects on the formation and evolution of galaxies.
}
   {The goal of this paper is to provide a measure of the environment of the isolated galaxies in the AMIGA sample, quantifying the influence of the candidate neighbours identified in our previous work and their potential effects on the evolution of the primary galaxies. Here we provide a quantification of the isolation degree of the galaxies in this sample.
}
   {Our starting sample is the Catalogue of Isolated Galaxies (CIG). We used two parameters to estimate the influence exerted by the neighbour galaxies on the CIG galaxy: the local number density of neighbour galaxies and the tidal strength affecting the CIG galaxy. We show that both parameters together provide a comprehensive picture of the environment. For comparison, those parameters have also been derived for galaxies in denser environments such as triplets, groups and clusters.
}
   {The CIG galaxies show a continuous spectrum of isolation, as quantified by the two parameters, from very isolated to interacting. The fraction of CIG galaxies whose properties are expected to be influenced by the environment is however low (159 out of 950 galaxies). The isolated parameters derived for the comparsion samples gave higher values than for the CIG and we found clear differences for the average values of the 4 samples considered, proving the sensitivity of these parameters.
}
   {The environment of the galaxies in the CIG has been characterised, using two complementary parameters quantifying the isolation degree, the local number density of the neighbour galaxies and the tidal forces affecting the isolated galaxies. A final catalogue of galaxies has been produced and the most isolated of these galaxies are consequently appropriate to serve as a reference sample for the AMIGA project.
}


   \maketitle
%
\section{Introduction} \label{sec:intro}

Although it is now generally recognised that the environment experienced by the galaxies during their whole lifetime plays a role as important as the initial conditions of their formation, there are still many open questions. In order to define what is the amplitude and dispersion of a given galaxy property that can be ascribed to ``nature'', a well characterised reference sample of isolated galaxies is needed.

The AMIGA project ({\bf A}nalysis of the interstellar {\bf M}edium of {\bf I}solated {\bf GA}laxies\footnote{{\tt http://www.iaa.es/AMIGA.html}}) aims to build and parametrise a statistically significant control sample of the most isolated galaxies in the local Universe. Our goal is to quantify the properties of different phases of the interstellar media of this sample. In an earlier paper \citep{2003A&A...411..391L}, we systematically revised all the positions of the galaxies in the Catalogue of Isolated Galaxies \citep[CIG,][]{1973AISAO...8....3K} in order to provide new values suitable for accurate telescope pointings or cross-correlations with on-line databases. The whole CIG was optically characterised in \citet{2005A&A...436..443V}, where we produced an optical luminosity function for all the galaxies. The physical distribution of the CIG galaxies with respect to the local large scale structure was also investigated. As expected we see little correspondence between the positions of the nearby cluster cores and the CIG galaxies, but some correspondence with the more complex local large-scale structure components has been found \citep{1983ApJ...275..472H}. The CIG redshift distribution re-enforces the evidence for a bimodal structure: peaks near 1500 and 6000~\kms\ correspond respectively to galaxies in the local supercluster and those in more distant large-scale components (particularly Perseus-Pisces). These two peaks in the redshift distribution are superimposed on a more homogeneous distribution involving about 50\% of the CIG \citep{2005A&A...436..443V}. Hence, the CIG 2D distribution is reasonably homogeneous as we would expect for a distribution sampling, predominantly, the preriferies of large-scale structures \citep{1981A&A....97..223B}. In 3D, the distribution is affected by the local and Pisces-Perseus superclusters. In particular, while the CIG likely contains many of the most isolated galaxies in the local Universe, it is not biased for galaxies in voids because we are usually looking through the front side of the bubble of galaxies sourrounding the void. Thus void galaxies often fail the isolation requirement. Additionally, the morphologies of the CIG galaxies were revised and type-specific optical luminosity functions derived in \citet{2006A&A...449..937S}. Mid- and far-infrared basic properties have been also derived for the CIG using data from the IRAS survey \citep{2006astro.ph.10784L}.

Studies of radio continuum, atomic and molecular gas \citep{2005A&A...442..455E,2006.PhD.Thesis.E}, CO and H$\alpha$ emission properties \citep{2005.PhD.Thesis.V} are in progress as well as a study of the small AGN population found in the sample. In these works we have identified several cases of CIG galaxies where \citeauthor{1973AISAO...8....3K}'s isolation criterion was failed. This motivated us to perform a careful reevaluation of the degree of isolation of the CIG, which was presented in \citet{2007A&A...1..1V} (hereafter, AMIGA\,IV). There we revised the environment of all the 950 CIG galaxies with radial velocities larger than 1500\kms\ in a minimum physical radius of 0.5~Mpc. We made use of POSS-I digitised plates on which we SExtracted \citep{1996A&AS..117..393B} all the objects brighter than $M_B \approx 17.5$ around the primary CIG galaxies. We used a version of the LMORPHO software \citep{1995PASP..107..770O,1996ApJ...472L..13O,2002ApJ...568..539O} adapted to our specific needs to separate stars from galaxies in order to produce the final catalogues of perturber galaxy candidates. Two visual checks were undertaken: the first one consisted in revising the types (galaxy, star, unknown) of all the extracted objects, by means of a Graphical User Interface displaying the types of the objects on the digitised POSS-I plates. The second visual check involved all the objects classified as galaxy for which the better resolution and dynamic range of the POSS-II allowed us to separate compact galaxies from stars. The final catalogue produced includes about 54\,000 neighbour galaxies containing the right ascension, declination, area, magnitude, diameter and projected separation to the associated CIG galaxy. For each galaxy, the area is given by the SExtractor software (converted to arcsec.$^2$), and the magnitude by the parameter MAG\_ISO. The velocities are heliocentric and we assume $H_{\rm 0} = 75$~\kms\ Mpc$^{-1}$.

Redshifts could be compiled for $\sim$ 30\% of the sample (16\,126 objects) from different catalogues in the bibliography and we were able to identify some galaxies failing \citeauthor{1973AISAO...8....3K}'s criterion with our new data. The presence of candidate neighbours, in a different number and with different sizes and magnitudes in the environments of CIG galaxies, leads us to go a step further. We provide a quantification of the degree of isolation of CIG galaxies according to different and complementary parameters, that will produce a well characterised picture of their environment.

The article is organised as follows: in Sect.~\ref{sec:param} we define the parameters that we will use to measure the isolation degree; in Sect.~\ref{sec:compa} these parameters are calculated for denser samples (triplets, groups and clusters of galaxies) for comparison. The results are presented in Sect.~\ref{sec:res}, and discussed in Sect.~\ref{sec:disc}, including the complementarity between the isolation parameters, the relationship with the \citeauthor{1973AISAO...8....3K}'s criterion, and finally the comparison with  similarly calculated isolation parameters for galaxies in denser environments. We also present our final catalogue of isolated galaxies. Section~\ref{sec:conc} summarises our work and presents the conclusions of our study.

\section{Quantification of the isolation} \label{sec:param}

We have used two complementary parameters in order to quantify the degree of isolation of the CIG galaxies: the local number density of neighbour galaxies and the tidal strength that these latter produce on the candidate isolated galaxy. Both parameters were calculated for all the 950 CIG galaxies considered in AMIGA\,IV (when excluding the 100 nearby galaxies with $V$ $<$ 1500\kms).

\subsection{Local number density} \label{sub:kth}

The local number density of the neighbour galaxies is calculated by focusing on the vicinity of the isolated galaxy candidates, where the principal perturbers should lie. An estimation of the local number density, $\eta_k$, is found by considering the distance to the $k^{\rm th}$ nearest neighbour. An unbiased estimator can be obtained if neither the central galaxy nor the $k^{\rm th}$ neighbour are counted \citep[see,][]{1985ApJ...298...80C, 2004MNRAS.349.1251M}. For this parameter, to minimise the contamination by background galaxies, only the neighbours with similar size \citep[0.25 to 4 times the diameter of each CIG galaxy, as defined by][]{1973AISAO...8....3K} are taken into account. To probe a local region around the central galaxy, we consider $k$ equal to 5, or less if there are not enough neighbours in the field:
\[
\eta_k \propto \log\bigg(\frac{k-1}{V(r_k)}\bigg)
\]
with $V(r_k) = 4 \pi r_k^3 /3$, where $r_k$ (in \arcm) is the projected distance to the $k^{\rm th}$ nearest neighbour.

The farther the $k^{\rm th}$ nearest neighbour, the smaller the local number density $\eta_k$. Hence this parameter provides a good description of the environment in the vicinity of the primary galaxy but with the disadvantage of not taking into account the mass (or size) of the perturbers.

This parameter could not be calculated for the full sample, since two neighbours is the necessary minimum and forty of the CIG galaxies in our sample did not fulfil this requirement.



\subsection{Tidal strength} \label{sub:TS}

In order to provide an estimation of the degree of isolation taking also into account the masses of the neighbours, we calculated the tidal strength affecting the primary CIG galaxies. To estimate the total tidal strength affecting each CIG galaxy (with a diameter $D_p$ and a mass $M_p$), we used a formalism  developed by \citet{1984AJ.....89..966D} to estimate the tidal strength affecting an extended object ($\Delta R$ is the extension of the object). The tidal strength per unit mass produced by a neighbour is proportional to $M_i R_{ip}^{-3}$, where $M_i$ is the mass of the neighbour, and $R_{ip}$ is its distance from the centre of the primary. However, no information on either $M_i$ or on the absolute $R_{ip}$ is available in most cases. We approximated $R_{ip}$ by the projected separation, $S_{ip}$, at the distance of the CIG galaxy. The dependence of $M_i$ on size, $M_i \propto D_i^\gamma$, is not well known, and we adopted $\gamma = 1.5$ \citep{1982ApJ...261..439R, 1984AJ.....89..966D}. Accordingly, the estimator $Q_{ip}$ is defined as the ratio between the tidal force and binding force. For one neighbour, the estimation of the interaction strength, reads:

\[ F_{\rm tidal} = \frac{M_i \times \Delta R}{R_{ip}^3} \simeq \frac{M_i \times D_p}{S_{ip}^3} \]
\[ F_{\rm bind} = \frac{M_p}{D_p^2} \]
\[Q_{ip} \equiv \frac{F_{\rm tidal}}{F_{\rm bind}} \propto \bigg(\frac{M_i}{M_p}\bigg) \bigg(\frac{D_p}{S_{ip}}\bigg)^3 \propto \frac{(\sqrt{D_p D_i})^3}{S_{ip}^3}\]

The logarithm of the sum of the tidal strength created by all the neighbours in the field, $Q = \log(\sum_i Q_{ip})$, is a dimensionless estimation of the gravitational interaction strength. The greater the value, the less isolated from external influence the galaxy, and vice-versa. A value of 0 indicates that the internal forces have the same amplitude as the sum of the tidal forces affecting the primary galaxy.

In spite of the lack of  redshift information for the projected neighbour galaxies (only 30\% with redshift, see AMIGA\,IV), $Q$ is expected to give a reasonable estimate of the tidal interaction strength in a statistical sense: if the candidate neighbour is a background object the true distance would have been underestimated, but also the true size and mass, hence both effects partly cancel out. Only in the case of a foreground object $Q$ is overestimated, but this effect is very marginal (see AMIGA\,IV, Fig. 8, where the foreground objects are statistically represented by the very small amount of neighbours having a negative magnitude difference with respect to the CIG galaxies).

The tidal strength ($Q$) has been calculated for all neighbours in the whole available fields searched (see AMIGA\,IV). In order to remove objects with the highest probabilities to be background and foreground neighbours, we also calculated the tidal strength ($Q_{Kar}$) taking into account only the similar size neighbours (factor 4 in size, as defined by \citeauthor{1973AISAO...8....3K}). For the 888 CIG galaxies with known redshifts, we also derived an estimation of the tidal strength ($Q_{0.5}$) produced by the neighbour galaxies lying within a physical radius of 0.5~Mpc from their associated CIG galaxy. This latter parameter was also calculated taking into account only the similar size neighbours lying within 0.5~Mpc ($Q_{0.5, Kar}$).

For the 62 CIG galaxies without known redshift, the tidal strength estimations were only calculated as produced by the neighbours in a square field of $ 55' \times 55'$ centred in each CIG galaxy ($Q$ and $Q_{Kar}$).

\begin{figure}
\includegraphics[width=1.1\columnwidth]{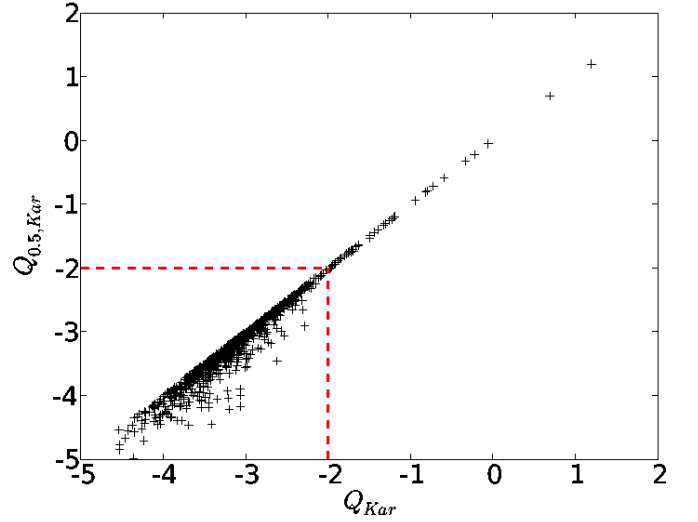}
\caption{Comparions of the tidal strength within a radius of 0.5~Mpc, $Q_{0.5}$, to the tidal strength in the whole available field, $Q$.}
\label{fieldSizeEffect}
\end{figure}

In order to evaluate the systematic errors that could be introduced by only considering this area, we have compared the tidal strength obtained for the galaxies with redshift considering a 0.5~Mpc radius and a $55' \times 55'$ field. The result is shown in Fig.~\ref{fieldSizeEffect} for the case where only similar size neighbours are taken into account (the result is similar when all neighbours are considered). The effect is marginal: CIG galaxies suffering the highest tidal strength ($Q_{Kar} \ge -2$) remain the same, and for the remaining galaxies ($Q_{Kar} < -2$), only a small trend is found: the value is slightly lower when only the neighbours within 0.5~Mpc are taken into account. This is due to the fact that adding new neighbour galaxies farther away than 0.5~Mpc has very little impact on the tidal strength affecting the CIG galaxy if this latter is not in a very low density environment. On the contrary, if the CIG galaxy had almost no neighbours within 0.5~Mpc, the addition of new neighbours outside this limit will enhance the estimation of the tidal strength, but with a small effect due to the large separation between the neighbours and the central CIG galaxy.

\section{Comparison samples} \label{sec:compa}

In order to compare the isolation degree of the CIG galaxies with galaxies in denser environments, we have selected three samples for comparison: triplets from the Karachentseva's catalogue \citep[KTG,][]{1979AISAO..11....3K}, compact groups from the Hickson catalogue \citep[HCG,][]{1982ApJ...255..382H} and Abell clusters \citep[ACO,][]{1958ApJS....3..211A,1989ApJS...70....1A}. The KTG and HCG catalogues complement the CIG since they were visually compiled using also an isolation criterion. For these samples, to avoid introducing any bias in the comparison, we followed the same reduction method described in detail in AMIGA\,IV (star/galaxy separation on the POSS images and visually inspection of the classification) and we calculated some of the isolation parameters ($\eta_k$, $Q$ and $Q_{Kar}$) as reviously obtained for the CIG, for fair comparisons.

\subsection{Karachentseva Triplets of Galaxies}

\citet{1979AISAO..11....3K} listed 84 northern isolated galaxy triplets compiled in a manner similar to the one used to compile the CIG. The apparent magnitudes are brighter than $m_{\rm Zw} = 15.7$ and the catalogue was built up on the basis of a complete examination of Palomar Sky Survey prints (POSS-I). \citet{1979AISAO..11....3K} showed that triple systems constitute 0.8\% of northern galaxies brighter than 15.7 mag, 64\% of the triplets are ``completely isolated'', and 24\% of the triplet members are elliptical and lenticular galaxies, while 76\% are spirals and irregulars.

In order to restrict our search for neighbour galaxies in $55' \times 55'$ fields (see AMIGA\,IV), we selected all triplets with the three galaxies having V $>$ 4687\kms. We applied the isolation parameters on the "A" galaxy (primary galaxy which will play the role of the CIG galaxy). This way, 41 triplets were selected (see Table~\ref{tab:ktg}). The coordinates (J2000), major axis and recession velocity of galaxy ``A'' are given.

\begin{table}
\caption{Studied subsample of Karachentseva Triplets of Galaxies$^\dag$.} \label{tab:ktg}
\begin{tabular}{c c c c c}
\hline \hline
KTG & RA (\degr) & Dec (\degr) & Major axis & Velocity\\
number & (J2000) & (J2000) & (\arcm) & (\kms)\\
\hline
 2 & 14.412716 & 43.800764 & 1.4 & 5539\\
 4 & 19.018667 & 46.730500 & 0.8 & 5602\\
 6 & 20.627667 & 39.199278 & 0.6 & 8084\\
 7 & 21.090833 & 32.224167 & 1.0 & 5214\\
10 & 48.980138 & 37.154116 & 0.6 & 6168\\
\vdots & \vdots & \vdots & \vdots & \vdots \\
\hline
\end{tabular}

$^\dag$The full table is available in electronic form at CDS or from {\tt http://www.iaa.es/AMIGA.html}.
\end{table}

\subsection{Hickson Compact Groups}


The Hickson Compact Group catalogue \citep[HCG,][]{1982ApJ...255..382H} is composed of 100 groups (largely quartets). Our selection process was the same as for the KTG, and we kept only the true physical groups following the work by \citet{1997ApJ...482..640S}. To fit in $55' \times 55'$ fields, the recession velocities had to be again greater than 4687\kms. A total of 34 Hickson Compact Groups remained. The coordinates (J2000), major axis and velocity are those of the galaxy on which the isolation parameters are applied, and are listed in Table~\ref{tab:hcg}.

\begin{table}
\caption{Studied subsample of Hickson Compact Groups$^\dag$.} \label{tab:hcg}
\begin{tabular}{c c c c c}
\hline \hline
HCG & RA (\degr) & Dec (\degr) & Major axis & Velocity\\
number & (J2000) & (J2000) & (\arcm) & (\kms)\\
\hline
 1 & 6.529708 & 25.725194 & 1.25 & 10237\\
 8 & 12.392292 & 23.578250 & 0.9 & 16077\\
10 & 21.590750 & 34.703028 & 3.0 & 5189\\
15 & 31.971167 & 2.167611 & 1.1 & 6967\\
17 & 33.52135 & 13.31104 & 0.36 & 18228\\
\vdots & \vdots & \vdots & \vdots & \vdots \\
\hline
\end{tabular}

$^\dag$The full table is available in electronic form at CDS or from {\tt http://www.iaa.es/AMIGA.html}.
\end{table}

\subsection{Abell clusters}

Only in the northern hemisphere, the Abell Clusters of Galaxies catalogue \citep{1958ApJS....3..211A,1989ApJS...70....1A} lists more than 2700 clusters classified in six richness classes (with only one cluster composing the richest class). We selected all clusters with available recession velocities between 4687 and 15\,000\kms. The ACO is a deeper sample than the CIG, KTG and HCG samples: the higher cut (15\,000\kms) is used in order to sample a volume of space roughly equivalent to the one spanned by the CIG (see Fig.~1 in AMIGA\,IV) and avoid possible biases. Among the clusters fulfilling these conditions, we selected the 15 clusters having a known diameter less than 55\arcm. This last condition ensures that we consider all the other galaxies of the cluster as neighbour galaxies interacting with the central primary galaxy. Table~\ref{tab:aco} summarises the main properties of the selected clusters (left) along with information on the primary galaxies (right) on which the isolation parameters have been applied (central cD galaxy, or brightest central galaxy, BCG).

\begin{table*}
\begin{center}
\caption[Abell clusters sample.]{ACO clusters (left) and primary galaxies selected (right).} \label{tab:aco}
\begin{tabular}{c c c c c c || c c c c c c }
\hline \hline
ACO & RA (h \farcm) & Dec (\degr $'$) & Velocity & Richness & Diameter & RA (\degr) & Dec (\degr) & Velocity & Major axis & Hubble\\
number & (J2000) & (J2000) & (\kms) & class & (\arcm) & (J2000) & (J2000) & (\kms) & (\arcm) & type\\
\hline
 160 & 01 12.9 & +15 31 & 13410 & 0 & 40 & 18.248726 & 15.491506 & 13137 & 0.78 & cD \\
 260 & 01 51.9 & +33 10 & 10440 & 1 & 50 & 28.024008 & 33.190811 & ... & 0.72 & ... \\
 671 & 08 28.5 & +30 25 & 14820 & 0 & 50 & 127.132118 & 30.432072 & 15087 & ... & ... \\
 957 & 10 14.0 & -00 55 & 13200 & 1 & 50 & 153.409729 & -0.925455 & 13293 & 1.5 & E+ pec \\
 999 & 10 23.4 & +12 51 & 9540 & 0 & 50 & 155.849396 & 12.835186 & 9764 & 1.2 & BCG \\
 1100 & 10 48.9 & +22 14 & 13650 & 0 & 40 & 162.190262 & 22.217989 & 13990 & 0.45 & BCG \\
 1177 & 11 09.5 & +21 42 & 9480 & 0 & 50 & 167.435104 & 21.759527 & 9589 & 1.8 & BCG \\
 1213 & 11 16.5 & +29 16 & 14040 & 1 & 50 & 169.095093 & 29.252588 & 13581 & 1.0 & SB0 \\
 2040 & 15 12.8 & +07 26 & 13680 & 1 & 32 & 228.197601 & 7.435083 & 13683 & 1.0 & BCG \\
 2152 & 16 05.4 & +16 27 & 11220 & 1 & 50 & 241.371292 & 16.435858 & 13211 & 1.3 & E \\
 2506 & 22 56.6 & +13 20 & 9930 & 1 & 20 & 344.288147 & 13.188705 & 6860 & 0.5 & ... \\
 2572 & 23 18.4 & +18 44 & 11850 & 0 & 50 & 349.626160 & 18.689167 & 11263 & 0.9 & cD \\
 2593 & 23 24.5 & +14 38 & 12990 & 0 & 50 & 351.084259 & 14.646864 & 12489 & 1.3 & cD \\
 2657 & 23 44.9 & +09 09 & 12420 & 1 & 46 & 356.239227 & 9.193000 & 12063 & 0.8 & cD \\
 2666 & 23 50.9 & +27 09 & 7950 & 0 & 50 & 357.744812 & 27.147602 & 8191 & 1.6 & cD \\
\hline
\end{tabular}
\end{center}
\end{table*}

\section{Results} \label{sec:res}

The values of the isolation parameters are listed in Table~\ref{tab:crit} which contains the following columns for each of the 950 CIG galaxies:
\begin{itemize}
\item {\it Column 1}: CIG number (from \citeauthor{1973AISAO...8....3K}'s original catalogue);
\item {\it Column 2}: $\eta_k$, local number density of similar size neighbours (arbitrary units);
\item {\it Column 3}: $k$, number of similar size neighbours taken into account to calculate $\eta_k$;
\item {\it Column 4}: $Q$, tidal strength estimation in the whole available square field;
\item {\it Column 5}: $Q_{Kar}$, tidal strength estimation of similar size neighbours in the whole available square field;
\item {\it Column 6}: $Q_{0.5}$, tidal strength estimation within 0.5~Mpc;
\item {\it Column 7}: $Q_{0.5,Kar}$, tidal strength estimation of similar size neighbours within 0.5~Mpc.
\end{itemize}

In the second and third columns, $\eta_k$ and $k$ are flagged with a value of ``-99.000'' when there are not, at least, two similar size neighbours. The parameter $k$ is equal to 5 for 835 CIG galaxies, equal to 4 for 24 galaxies, to 3 for 20 galaxies and to 2 for 31 galaxies. It is flagged with the value of ``-99.000'' for forty galaxies. The parameters involving the redshift of the CIG galaxies ($Q_{0.5}$ and $Q_{0.5, Kar}$ in Cols.~6 and 7, respectively) are arbitrarily equal to ``-98.000'' in Table~\ref{tab:crit} for the 62 galaxies with unknown redshifts (because there was no possibility to derive the physical radius of 0.5~Mpc).

\begin{table}
\caption{Isolation parameters calculated for the galaxies in the CIG$^\dag$.}
\label{tab:crit}
\begin{tabular}{c c c c c c c}
\hline \hline
(1)  &  (2)  &(3)&  (4)  &  (5)  &  (6)  &  (7)\\
CIG &$\eta_k$&$k$&  $Q$  & $Q_{Kar}$ & $Q_{0.5}$ & $Q_{0.5, Kar}$\\
\hline
  1 & 1.814 & 5 & -1.704 & -2.787 & -1.733 & -2.911\\
  2 & 0.971 & 5 & -3.565 & -3.565 & -3.936 & -3.936\\
  3 & 1.018 & 5 & -3.214 & -3.214 & -98.000 & -98.000\\
  4 & 0.987 & 4 & -2.050 & -3.736 & -2.059 & -3.797\\
  5 & 1.588 & 5 & -2.933 & -2.933 & -2.962 & -2.962\\
\vdots & \vdots & \vdots & \vdots & \vdots & \vdots & \vdots \\
\hline
\end{tabular}

$^\dag$The full table is available in electronic form at CDS or from {\tt http://www.iaa.es/AMIGA.html}.
\end{table}

The values of the isolation parameters ($\eta_k$ along with $k$, $Q$ and $Q_{Kar}$) for the denser samples are listed in Tables~\ref{tab:critKTG}, \ref{tab:critHCG} and \ref{tab:critACO}, for the KTG, HCG and ACO samples, respectively.

\begin{table}
\caption{Isolation parameters calculated for the galaxies in the KTG$^\dag$.}
\label{tab:critKTG}
\begin{tabular}{c c c c c}
\hline \hline
(1)  &  (2)  &(3)&  (4)  &  (5)\\
KTG &$\eta_k$&$k$&  $Q$  & $Q_{Kar}$\\
\hline
 2 & 3.080 & 5 & 1.658 & 1.658\\
 4 & 1.842 & 5 & -0.960 & -1.007\\
 6 & 1.513 & 5 & -2.428 & -2.428\\
 7 & 2.271 & 5 & -1.862 & -2.051\\
10 & 1.342 & 5 & -1.864 & -1.864\\
\vdots & \vdots & \vdots & \vdots & \vdots\\
\hline
\end{tabular}

$^\dag$The full table is available in electronic form at CDS or from {\tt http://www.iaa.es/AMIGA.html}.
\end{table}

\begin{table}
\caption{Isolation parameters calculated for the galaxies in the HCG$^\dag$.}
\label{tab:critHCG}
\begin{tabular}{c c c c c}
\hline \hline
(1)  &  (2)  &(3)&  (4)  &  (5)\\
HCG &$\eta_k$&$k$&  $Q$  & $Q_{Kar}$\\
\hline
 1 & 2.504 & 5 & 0.853 & 0.850\\
 8 & 3.175 & 5 & 0.555 & 0.553\\
 10 & 2.569 & 5 & -0.096 & -0.099\\
 15 & 3.184 & 5 & -1.242 & -1.249\\
 17 & 1.616 & 5 & 0.295 & 0.295\\
\vdots & \vdots & \vdots & \vdots & \vdots\\
\hline
\end{tabular}

$^\dag$The full table is available in electronic form at CDS or from {\tt http://www.iaa.es/AMIGA.html}.
\end{table}

\begin{table}
\caption{Isolation parameters calculated for the galaxies in the ACO sample.}
\label{tab:critACO}
\begin{tabular}{c c c c c}
\hline \hline
(1)  &  (2)  &(3)&  (4)  &  (5)\\
ACO &$\eta_k$&$k$&  $Q$  & $Q_{Kar}$\\
\hline
 160 & 3.797 & 5 & -1.014 & -1.157\\
 260 & 3.044 & 5 & -1.706 & -1.706\\
 671 & 3.953 & 5 & -0.710 & -0.757\\
 957 & 3.763 & 5 &  0.410 & -0.663\\
 999 & 4.018 & 5 & -0.174 & -0.193\\
1100 & 2.837 & 5 & -1.056 & -1.058\\
1177 & 2.588 & 5 &  0.541 & -0.503\\
1213 & 4.415 & 5 & -0.224 & -0.225\\
2040 & 3.338 & 5 & -0.027 & -0.031\\
2152 & 3.718 & 5 &  0.349 &  0.340\\
2506 & 1.818 & 5 & -1.211 & -1.211\\
2572 & 2.971 & 5 & -1.217 & -1.505\\
2593 & 3.739 & 5 &  0.167 & -0.996\\
2657 & 3.438 & 5 &  0.174 &  0.168\\
2666 & 4.020 & 5 &  0.001 & -0.161\\
\hline
\end{tabular}
\end{table}

\begin{figure}
\includegraphics[width=1.1\columnwidth]{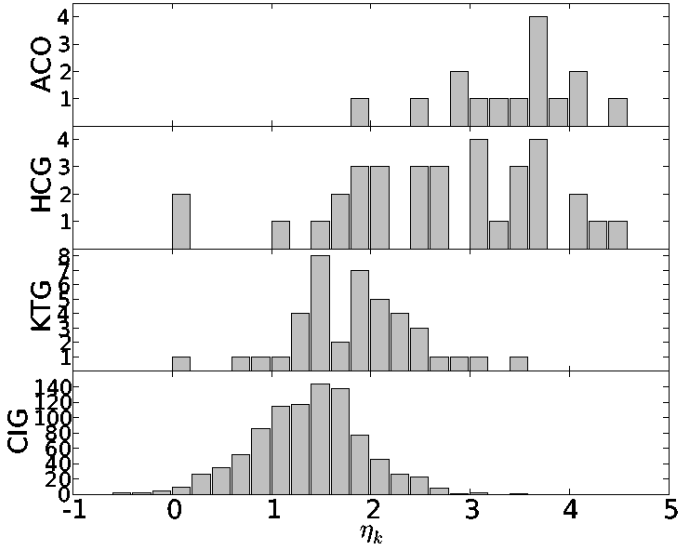}
\caption{Distribution of the local number density ($\eta_k$) estimation, for the four samples of galaxies.} \label{fig:hist_LND_4}
\end{figure}

\begin{figure}
\includegraphics[width=1.1\columnwidth]{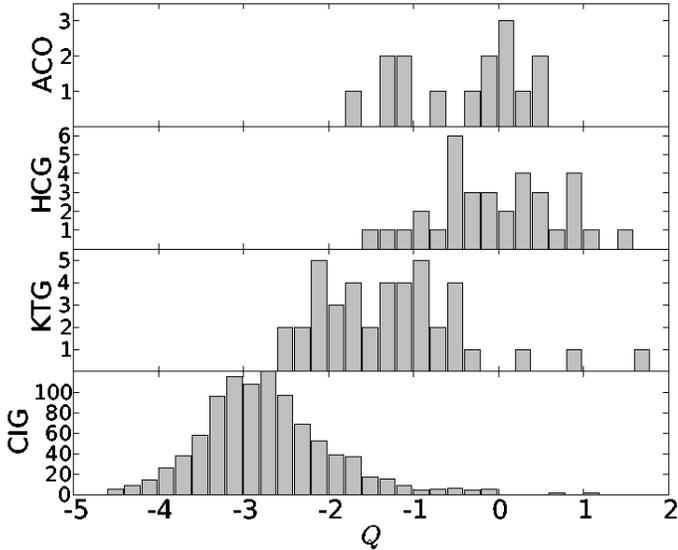}
\caption{Distribution of the tidal strength ($Q$) estimation including all neighbours, for the four samples of galaxies.} \label{fig:hist_TSall_4}
\end{figure}

\begin{figure}
\includegraphics[width=1.1\columnwidth]{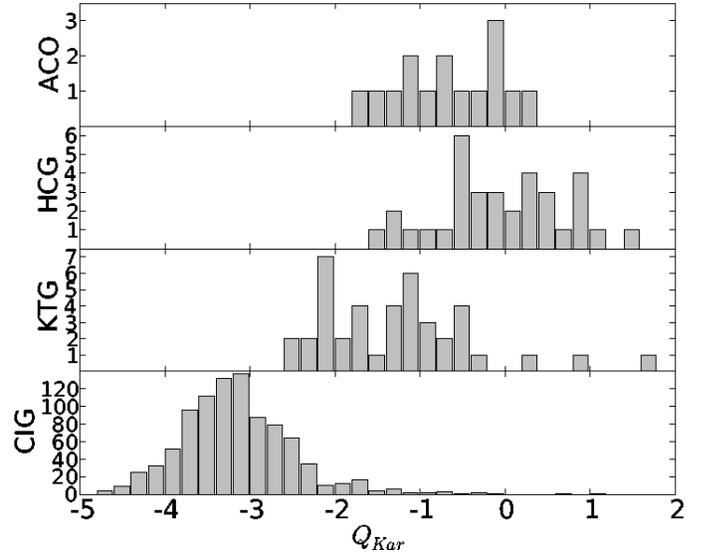}
\caption{Distribution of the tidal strength ($Q_{Kar}$) estimation including similar size neighbours, for the four samples of galaxies.} \label{fig:hist_TSkar_4}
\end{figure}

In Figs.~\ref{fig:hist_LND_4}--\ref{fig:hist_TSkar_4}, the histograms of the isolation parameters are plotted, for the four samples. As expected the vast majority of CIG galaxies are lying in less dense environments compared to the KTG, HCG and ACO samples. Only a small fraction of the CIG galaxies are lying in an environment that can affect their evolution as much as for galaxies in the comparsion sample.

The trend of the mean values from one sample to another shows that the isolation parameters are sensitive enough to the addition of one neighbour in the vicinity of the primary galaxy: the triplets and compact groups (mainly four galaxies) always show values well separated for both the local number density and the tidal strength estimations (see Table~\ref{tab:stat}).

\begin{table}
\caption{Means and standard deviations of the isolation parameters for the CIG and comparison samples.}
\label{tab:stat}
\begin{tabular}{l c c c c}
\hline \hline
               &  CIG  &  KTG  &  HCG  &  ACO\\
\hline
mean($\eta_k$) & 1.378 & 1.863 & 2.722 & 3.430\\
std($\eta_k$)  & 0.556 & 0.629 & 1.105 & 0.676\\
mean($Q$)      & -2.720 & -1.222 & -0.026 & -0.380\\
std($Q$)       & 0.760 & 0.866 & 0.726 & 0.710\\
mean($Q_{Kar}$)& -3.142 & -1.251 & -0.034 & -0.644\\
std($Q_{Kar}$) & 0.689 & 0.884 & 0.738 & 0.621\\
\hline
\end{tabular}
\end{table}

For the tidal strengths, a value of $-2$ is a key value \citep{1984PhR...114..321A}: it represents an external influence amounting to 1\% of the internal forces. Likewise it corresponds to a threshold separating the galaxies affected by their environment from galaxies evolving without external perturbations. Basically it separates the interactions which will affect a galaxy or not. Comparing the tidal forces created by all the neighbours ($Q$, Fig.~\ref{fig:hist_TSall_4}) with the ones created only by the similar size neighbours ($Q_{Kar}$, Fig.~\ref{fig:hist_TSkar_4}), we can see that the influence of the small neighbours is not negligible. In some cases, the influence of the small neighbours can increase the tidal strength and the value can reach the threshold of $-2$: hence, the small neighbours very near the isolated galaxies can highly influence their evolution.

\section{Discussion} \label{sec:disc}

\subsection{Complementarity between the isolation parameters}

\begin{figure}
\includegraphics[width=1.1\columnwidth]{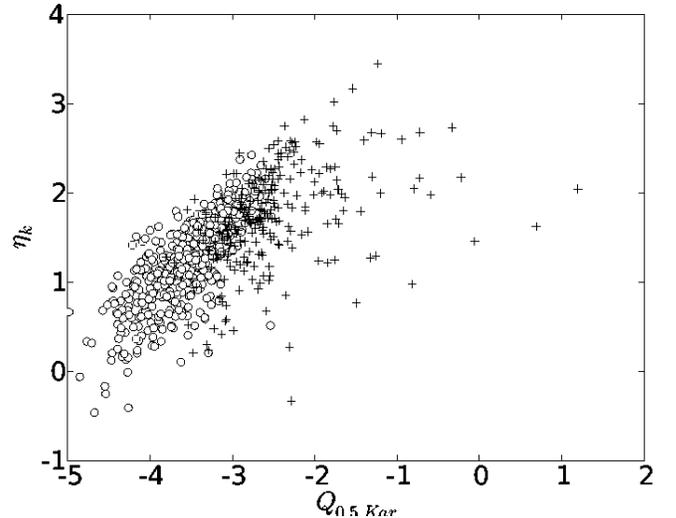}
\caption{Local number density vs. tidal strength ($Q_{0.5, Kar}$) estimations for similar size neighbours within 0.5~Mpc from the CIG galaxy. Galaxies failing are depicted by pluses and galaxies in agreement with the \citeauthor{1973AISAO...8....3K}'s criterion are displayed with circles.} \label{fig:kara01}
\end{figure}

The two isolation parameters give consistent results, as shown in Fig.~\ref{fig:kara01} (local number density vs. tidal strength). When a galaxy shows low values for both the local number density and the tidal strength estimation, this galaxy is very isolated from any sort of external influence. On the contrary, when the two values are high, the evolution of the galaxy can be perturbed by the environment and this kind of galaxy is not suitable to represent the normal features of isolated galaxies.

The two isolation parameters are also complementary between each other. For instance, a neighbour very close to a CIG galaxy would be counted as one object by the local number density estimation but will drastically increase the value of the tidal strength affecting the CIG galaxy (see for instance the points to the right part in Fig.~\ref{fig:kara01}). On the other hand, if the tidal strength is low but the local number density high, we can conclude that the environment consists of relatively small neighbours, present in a high number near the primary galaxy. This latter case excludes, for example, major interactions. The use of the combination of these various parameters allows us to have a clear picture of the environment around the candidate isolated galaxies.

To illustrate the complementarity between the isolation parameters, we can focus on three representative cases. CIG 918 has very low values of both the local number density ($\eta_k = -0.169$) and tidal strength estimation ($Q = -4.432$). Only 6 neighbours are identified in the surrounding field, all have a similar size to CIG 918, but none violates the \citeauthor{1973AISAO...8....3K}'s criterion. On the contrary, CIG 1030 possesses high values of the two isolation parameters ($\eta_k = 3.013$ and $Q = -1.741$), 131 neighbour galaxies are identified in the field, among which 4 are violating the \citeauthor{1973AISAO...8....3K}'s criterion. Last, with a relatively low local number density ($\eta_k = 0.976$) and a high tidal strength estimation ($Q = -0.813$), CIG 532 represents an intermediate case. Thirty neighbour galaxies were identified in the field, one of them very near (at a projected distance of 35\arcs) to CIG 532 and violating the \citeauthor{1973AISAO...8....3K}'s criterion. The presence of one single similar size neighbour at such a small distance to the CIG galaxy explains the high value of the tidal strength estimation while the local number density, averaged over the 5 nearest neighbours, remains quite low.


\subsection{Karachentseva's criterion as a function of the isolation parameters}

In Fig.~\ref{fig:kara01}, \citeauthor{1973AISAO...8....3K}'s criterion (following the study done in AMIGA\,IV) is displayed as a function of the local number density versus the tidal strength estimation: galaxies failing are depicted by pluses and galaxies in agreement with \citeauthor{1973AISAO...8....3K}'s criterion are displayed with circles. The two subsamples are well separated by the isolation parameters: as expected, the galaxies that are failing the \citeauthor{1973AISAO...8....3K}'s criterion represent a population more strongly interacting with their environment that the rest of the sample.

As seen in AMIGA\,IV, the \citeauthor{1973AISAO...8....3K}'s original criterion is very restrictive and the galaxies failing it but having an estimation of tidal strength minor than -2 are galaxies only marginally affected by their environment \citep{1984PhR...114..321A} and still suitable to represent isolated galaxies.

\subsection{Comparison with the denser environment samples}

We applied \citeauthor{1973AISAO...8....3K}'s criterion to the primary galaxies in the denser environment samples and all the primary galaxies in triplets, groups and clusters, as expected, fail it.

\begin{figure}
\includegraphics[width=1.1\columnwidth]{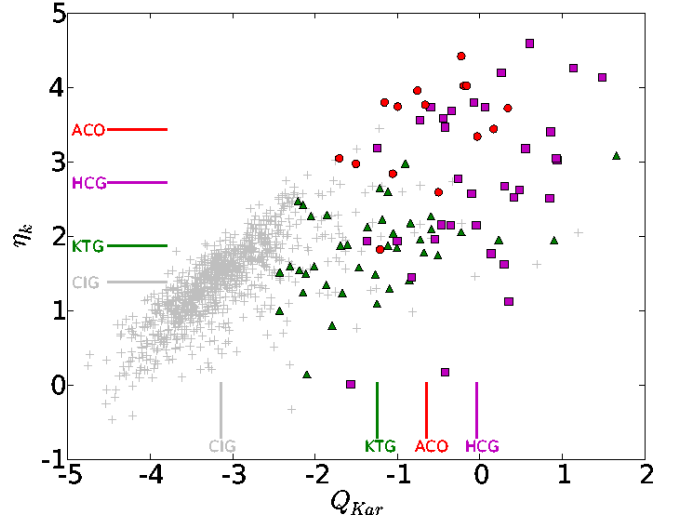}
\caption{Comparison between the local number density and tidal strength ($Q_{Kar}$) parameters for the CIG and the comparison samples. The CIG galaxies are represented by grey pluses. The KTG are depicted by green triangles, the HCG by magenta squares and the ACO by red dots. The mean values of each sample are shown by horizontal and vertical lines, following the same colour code.
} 
\label{fig:cigKtgHcgAco}
\end{figure}

Figure~\ref{fig:cigKtgHcgAco} shows the comparison between the local number density and the tidal strength estimation for the CIG and the other catalogues sampling denser environments. The first result is that the mean tidal strength estimation increases clearly from CIG to KTG and HCG samples. Hence it is sensible enough to distinguish between environments dominated by 1, 3 and 4 galaxies. Along the ordinate, reflecting the local number density estimation, there is some overlap between these three samples. This reflects the fact that the KTG and the HCG are samples constructed also with the help of isolation requirements : they are isolated triplets and isolated groups. But, by definition, two of the triplet galaxies and at least three of the group galaxies are very close to the primary galaxy. This is why the $5^{\rm th}$ neighbour is, on average, closer to the triplet or the group than the $5^{\rm th}$ neighbour of an isolated galaxy. The value of the local number density estimation is the result of a compromise between the two effects just cited. The CIG, KTG and HCG samples are also in a logical order along the ordinate, but less separated than along the abcissa. The tidal strength and the local number density estimations are complementary parameters and it is important to use both in order to have an accurate picture of the repartition of galaxies surrounding a primary galaxy.

The ACO clusters represent physically very different entities from the CIG, KTG or HCG catalogues as they can involve several thousands of galaxies. The ACO subsample considered in our study is biased towards the poorest clusters. The 15 clusters selected are not representative of the mean characteristics of the ACO sample for various reasons: they are among the nearest ones and belong to the two poorest richness classes. For technical reasons, they also had to possess a known diameter minor than 55\arcm. The average value of the tidal strength suffered by the primary ACO galaxies are in between the KTG and HCG ones. One may expect that result as the HCGs are the densest concentrations of galaxies in the Universe. On the other hand the ACO sample shows the highest local number density estimation, as expected, because all the 5 nearest neighbours are within the core of the clusters and there is no effect due to an isolation requirement (contrarily to the CIG, KTG and HCG samples).

\subsection{Comparison to visually identified interacting CIG galaxies}

In \citet{2006A&A...449..937S}, when visually inspecting the morphology of the CIG galaxies, a significant number of CIG galaxies were flagged for the presence of nearby neighbours or signs of distortion likely due to interaction. \citet{2006A&A...449..937S} list 32 CIG galaxies as members of interacting systems and also noted 161 galaxies where interaction is suspected based upon evidence for asymmetries/distortions that might be of tidal origin.

\begin{figure}
\includegraphics[width=1.1\columnwidth]{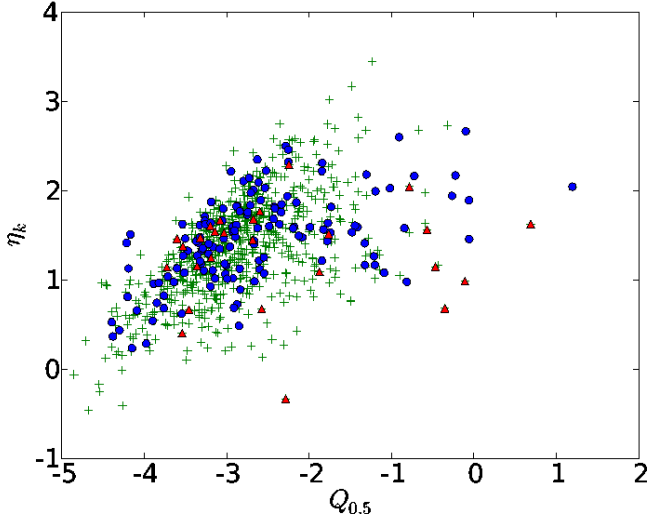}
\caption{The tidal strength ($Q_{0.5}$) is calculated including all the neighbours within 0.5~Mpc. The CIG galaxies, which were not flagged as interacting galaxies, are represented by green pluses. The red triangles indicate a morphologically distorted system and/or almost certain interacting system while the blue circles indicate evidence for interaction/asymmetry with/without certain detection of a neighbour \citep{2006A&A...449..937S}.} \label{fig:sule_y_bo}
\end{figure}

\begin{figure}
\includegraphics[width=1.1\columnwidth]{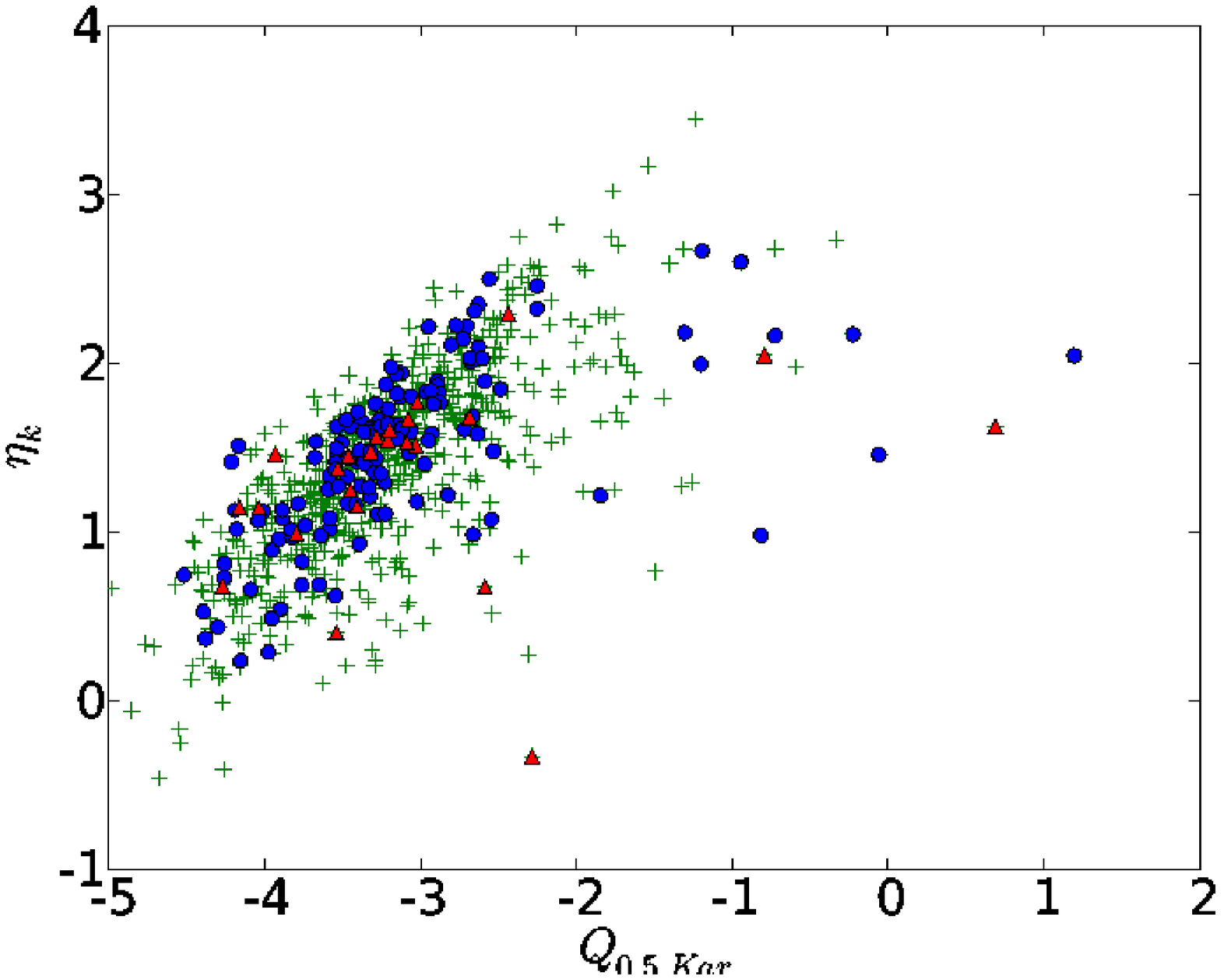}
\caption{The tidal strength ($Q_{0.5,Kar}$) is calculated including only the similar size neighbours within 0.5~Mpc. The CIG galaxies, which were not flagged as interacting galaxies, are represented by green pluses. The red triangles indicate a morphologically distorted system and/or almost certain interacting system while the blue circles indicate evidence for interaction/asymmetry with/without certain detection of a neighbour \citep{2006A&A...449..937S}.} \label{fig:sule_y_bo_sz}
\end{figure}

We checked the local number density and the tidal strength estimation for these two populations. They are over-plotted in Figs.~\ref{fig:sule_y_bo} (tidal strength computed including all the neighbour galaxies) and \ref{fig:sule_y_bo_sz} (tidal strength calculated including only the similar size neighbours). In Fig.~\ref{fig:sule_y_bo}, we see that the majority of the galaxies having high tidal strength estimation were also flagged by \citet{2006A&A...449..937S}. If we consider only the influence of the similar size neighbours (Fig.~\ref{fig:sule_y_bo_sz}), the effect is less clear.

\begin{figure}
\includegraphics[width=1.1\columnwidth]{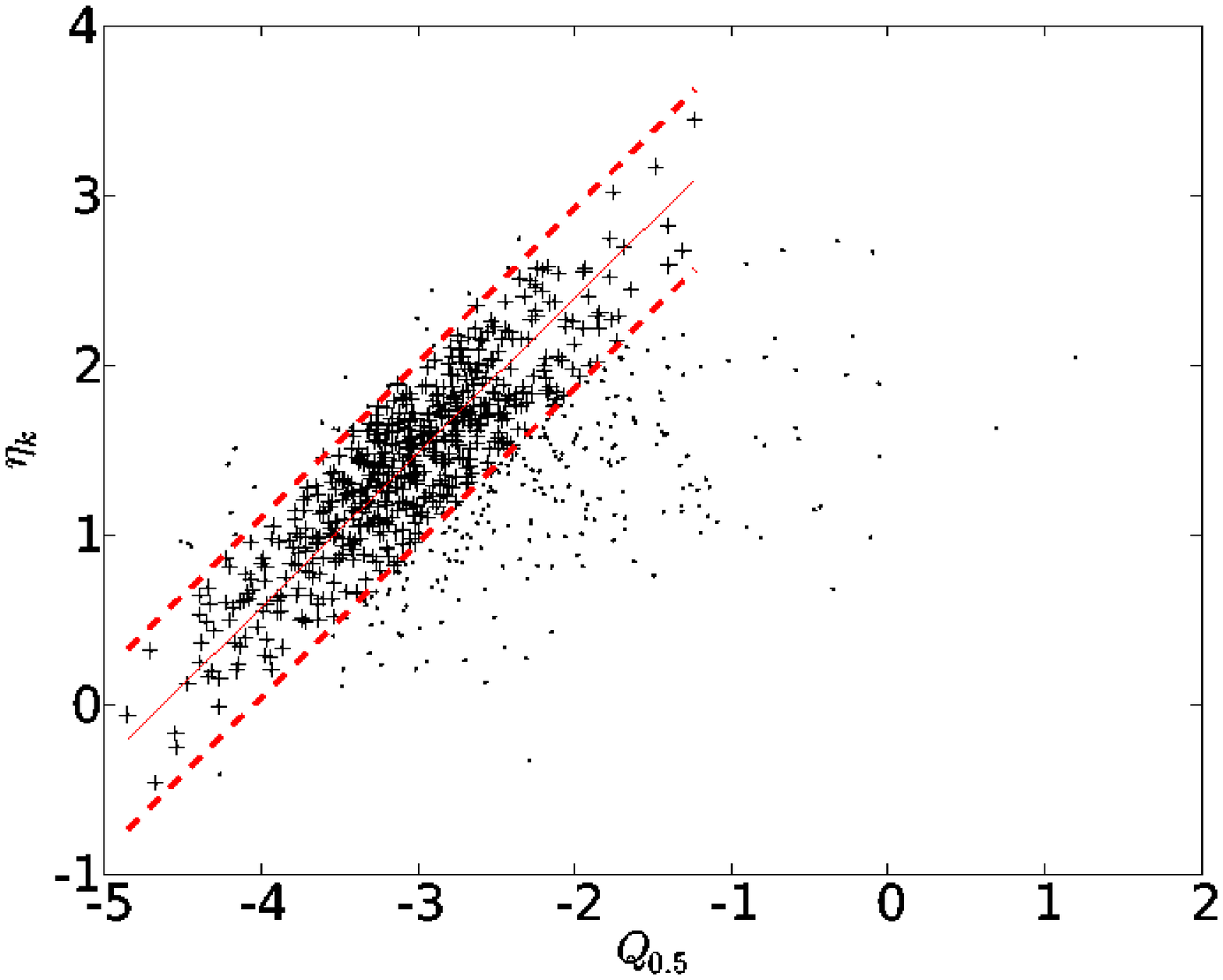}
\caption{Local number density vs. tidal strength ($Q_{0.5}$). The final bisector fit is shown by the plain red line. The dotted red lines represent the final 2 $\sigma$ dispersion: the CIG galaxies within these limits are depicted by pluses, the CIG rejected during the successive iterations are shown by points.}
\label{fig:fitDev}
\end{figure}

To quantify this effect, we made a regression analysis (with a 2$\sigma$ rejection limit), fitting a bisector to the local number density and tidal strength estimation $Q_{0.5}$. The convergence is reached after 26 iterations: Fig.~\ref{fig:fitDev} shows the bisector, the 2$\sigma$ limits and the CIG galaxies rejected. In this figure, the black pluses enclosed within the 2$\sigma$ limits represent the CIG galaxies having the environment the most representative of the CIG sample, while the black points depict galaxies in the CIG having peculiar environments with respect to the bulk of the galaxies composing the CIG. For each galaxy, we then calculated the orthogonal distance to the fitted bisector.

Optical asymmetries are often observed in galaxies: about 30\% of field galaxies exhibit significant lopsidedness, not necessarily due to a close, current interaction \citep{1997ApJ...477..118Z}; in some cases, the perturbative companion could have merged with the primary galaxy or could have receded too far away for the interaction to be still evident. \citet{1994A&A...290L...9R} found that the percentage of (fields and supercluster fields) galaxies showing asymmetries in the distribution of neutral hydrogen can even reach at least 50\% (the non-circular motions can be observed in the \ion{H}{I} maps or in the global profiles). Reciprocally, some galaxies show no visual disturbance but display kinematic evidence for interactions, such as counter-rotating disks: see for instance NGC~3593 \citep{1996ApJ...458L..67B} or NGC~4138 \citep{1996AJ....112..438J}. Hence, the remaining of the study only concern the 32 galaxies for which the presence of a nearby neighbour was clearly identified.

\begin{figure}
\includegraphics[width=1.1\columnwidth]{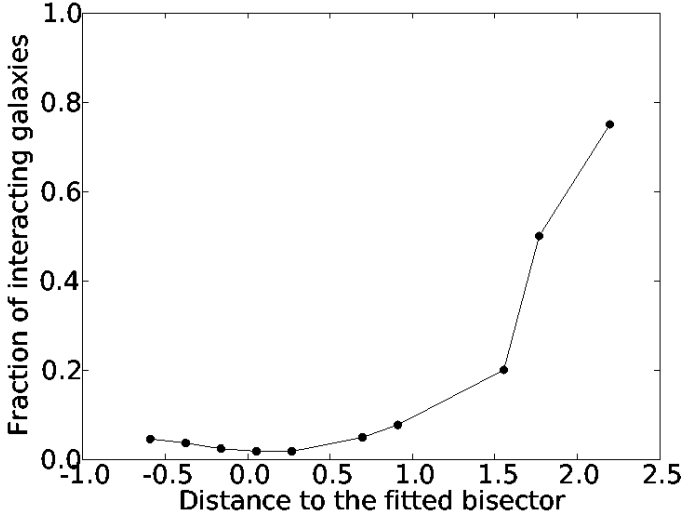}
\caption{The fraction of interacting galaxies is shown as a function of the distance to the fitted bisector (calculated with $Q_{0.5}$). As the environment of the galaxies gets denser and denser (large values for the distance, departure from the normal environments of the CIG galaxies), the galaxies have a higher probability to be disturbed due to interactions with their environment.}
\label{fig:frac_y}
\end{figure}

In Fig.~\ref{fig:frac_y}, we show the fraction of galaxies that were flagged as interacting galaxies as a function of the orthogonal distance to the bisector fit: it clearly appears that the farther a galaxy from the most common environment in the CIG, the higher its probability to show features driven by interactions.

\begin{figure}
\includegraphics[width=1.1\columnwidth]{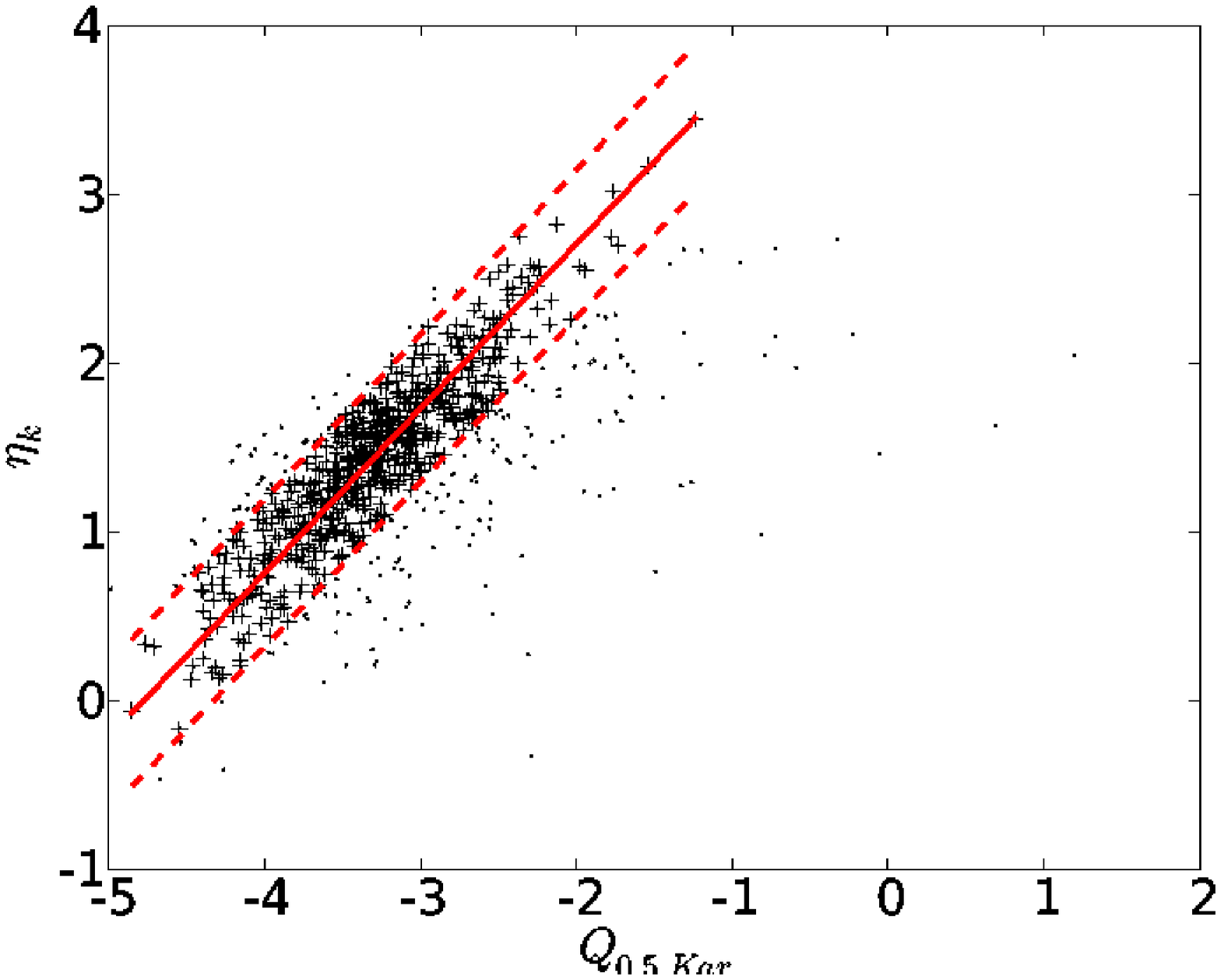}
\caption{Local number density vs. tidal strength ($Q_{0.5,Kar}$). The final bisector fit is shown by the plain red line. The dotted red lines represent the final 2 $\sigma$ dispersion: the CIG galaxies within these limits are depicted by pluses, the CIG rejected during the successive iterations are shown by points.}
\label{fig:fitDev_sz}
\end{figure}

\begin{figure}
\includegraphics[width=1.1\columnwidth]{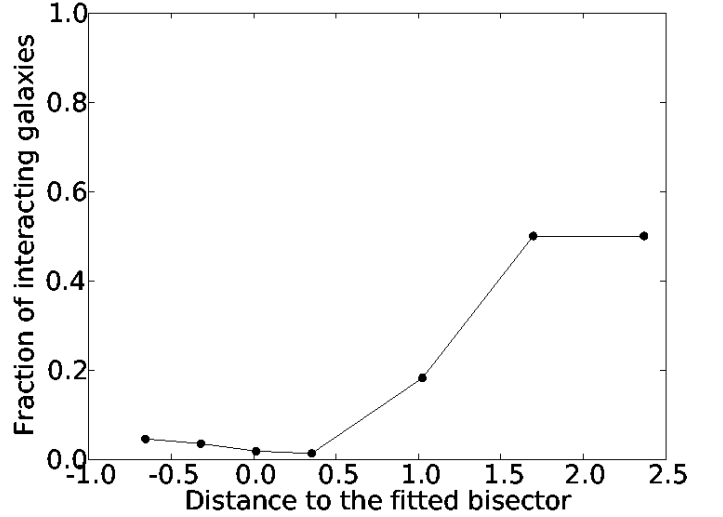}
\caption{The fraction of interacting galaxies is shown as a function of the distance to the fitted bisector (calculated only including the similar size neighbours within 0.5~Mpc of each CIG galaxy).}
\label{fig:frac_sz_y}
\end{figure}

The same study was done taking only into account the similar size neighbours to estimate the tidal strength $Q_{0.5,Kar}$. The best fit was reached by the regression analysis (with a 2$\sigma$ rejection limit) after 13 iterations, see Fig.~\ref{fig:fitDev_sz}. The fraction of galaxies that were optically flagged as interacting galaxies as a function of the orthogonal distance to the fitted bisector is shown Fig.~\ref{fig:frac_sz_y}. The fraction of the interacting galaxies shows the same trend than the study done with $Q_{0.5}$, but it obviously lacks to tackle some of the optically flagged galaxies (for a distance to the fitted bisector higher than $\sim$1.7). This difference between the two studies shows the importance to take into account the small neighbours which can produce major perturbations on the CIG galaxies.

\subsection{A revised catalogue of isolated AMIGA galaxies}

The main aim of this study was to produce a catalogue of isolated galaxies and to have, for each, one clear picture of its environment. Starting from the CIG, we will remove all the galaxies whose evolution could have been affected by their environment. We saw that perturbations could arise when the tidal forces affecting a galaxy amounted to at least 1\% of the internal binding forces \citep{1984PhR...114..321A, 2004A&A...420..873V}. For the local number density, this translates to a value of 2.4. This value is given by the final fitted bisector correlation, calculated including all the neighbours (even the small, not considered by \citeauthor{1973AISAO...8....3K} but for which \citet{2006A&A...449..937S} detected major, obvious role in the evolution of the galaxies). This two values characterise best the limits for selecting a sample of isolated galaxies, using the complementarity between the local number density and the tidal strength estimation. One hundred twenty five CIG galaxies are affected by tidal forces that could perturb their evolution ($Q \ge -2$), sixteen galaxies are lying in relatively high number density environment ($\eta_k \ge 2.4$) and eighteen galaxies cumulate the two conditions. Consequently, a total of 159 CIG galaxies are lying in environments that could affect their evolution and are not suited to be included in a sample of isolated galaxies. We remove these galaxies and the remaining 791 galaxies define the AMIGA sample of isolated galaxies. This is still a numerous enough sample allowing statistical significance, even for subsamples of galaxies (e.g.: sorted by morphologies). The 791 galaxies show a continuous gradient of interaction with their environment but all have their evolution dominated by their intrinsic properties. Further AMIGA studies will consider this sample as a reference and study the properties of the ISM of these galaxies in various wavelengths. The comparison with the properties of galaxies lying in denser environments or in interaction, will permit to quantify the effects added by the environment on the intrinsic evolution of galaxies.

\section{Summary and conclusions} \label{sec:conc}

All the work performed in previous papers as well as in the current work allows us to reach our original goal: to refine the CIG in order to provide the best possible sample of the most isolated galaxies, and have a quantification of the degree of isolation of each galaxy. We have performed the following refinements:
\begin{enumerate}
\item Using the $\sim$54\,000 neighbour galaxies listed in AMIGA\,IV, we calculated continuous parameters of isolation for 950 galaxies in the CIG. We used the local number density and the tidal force estimation to precisely describe the environment of each of the 950 CIG galaxies.
\item These two isolation parameters are complementarity and allow us to have a clear picture of the environment of each CIG galaxy considered.
\item We compared the level of isolation of the galaxies in the CIG with galaxies in denser environments: 41 triplets, 34 groups and 15 clusters were selected for the comparison. The two isolation parameters are very well suited to discriminate the isolated galaxies from galaxies more in interaction with their environments.
\item The galaxies flagged in AMIGA\,IV as failing the \citeauthor{1973AISAO...8....3K}'s original criterion also belong to the most interacting galaxies defined by the two isolation parameters.
\item The CIG galaxies flagged by \citet{2006A&A...449..937S} as optically distorted galaxies also show the highest values of the tidal strength estimation. The presence of small neighbours near the primary galaxy has a notable effect in the visual identification of distorted galaxies. Their contribution has to be taken into account in order to be sure to select galaxies isolated not only from similar size neighbours \citep[as done by][]{1973AISAO...8....3K} but also from dwarfs.
\item We give a final catalogue of 791 isolated galaxies, showing continuous isolation degrees, but with their evolution mainly driven by intrinsic properties and not by the external influence of their environment. These galaxies represent the basic AMIGA sample of isolated galaxies that will be considered as a reference for the further AMIGA studies, mapping the ISM in various wavelengths.
\end{enumerate}

\begin{acknowledgements}
S. V. thanks Gary A. Mamon, Chantal Balkowski, Alessandro Boselli, Santiago Garcia-Burillo, Jose M. Vilchez for their comments and J.~A. Jim\'enez Madrid for the Debian support. We thank the anonymous referee for useful comments. This research has made use of the NASA/IPAC Extragalactic Database (NED) which is operated by the Jet Propulsion Laboratory, California Institute of Technology, under contract with the National Aeronautics and Space Administration, and of the Lyon Extragalactic Database (LEDA). This work has been partially supported by DGI Grant AYA 2005-07516-C02-01 and the Junta de Andaluc\'{\i}a (Spain). UL acknowledges support by the research project ESP\,2004-06870-C02-02. Jack Sulentic is partially supported by a sabbatical grant SAB2004-01-04 of the Spanish Ministerio de Educaci\'on y Ciencias. GB acknowledges support at the IAA/CSIC by an I3P contract (I3P-PC2005F) funded by the European Social Fund.
\end{acknowledgements}

\bibliography{7481ref}

\end{document}